\documentclass[11pt]{article}
\usepackage{setspace}
\usepackage{amsfonts}
\usepackage{amsmath, amsthm, amssymb}
\usepackage{amscd}
\usepackage{pictexwd,dcpic}

\newtheorem{definition}{Definition}[section]

\begin{document}

\title{A Formalism for Quantum Games and an Application}
\author{Steven A. Bleiler\footnote{Department of Mathematics and Statistics, Portland State University, PO Box 751, Portland, Oregon 97207-0751 USA. Email: bleilers@pdx.edu.}}
\maketitle
	
\begin{abstract}
This paper presents a new mathematical formalism that describes the quantization of games. The study of so-called quantum games is quite new, arising from a seminal paper of D. Meyer \cite{Meyer} published in Physics Review Letters in 1999. The ensuing near decade has seen an explosion of contributions and controversy over what exactly a quantized game really is and if there is indeed anything new for game theory. What has clouded many of the issues is the lack of a mathematical formalism for the subject in which these various issues can be clearly and precisely expressed, and which provides a context in which to present their resolution. Such a formalism is presented here, along with proposed resolutions to some of the issues discussed in the literature. One in particular, the question of whether there can exist equilibria in a quantized version of a game that do not correspond to classical correlated equilibria of that game and also deliver better payoffs than the classical correlated equilibria is answered in the affirmative for the Prisoner's Dilemma and Simplified Poker. 
\end{abstract}

\section{Classical Background}
As with all mathematical formalisms, it's best to begin with a definition.
\begin{definition}
Given a set $\{ 1, 2, \cdots, n \}$ of players, for each player a set $S_i$ $(i=1, \cdots, n)$ of so-called \emph{pure strategies}, and a set $\Omega_i$ $(i=1, \cdots, n)$ of \emph{possible outcomes}, a \emph{game} $G$ is a vector-valued function whose domain is the Cartesian product of the $S_i$'s and whose range is the Cartesian product of the $\Omega_i$'s. In symbols 
                       \[\prod_{i=1}^n S_i \stackrel{G}{\longrightarrow} \prod_{i=1}^n \Omega_i \]
The function $G$ is sometimes referred to as the \emph{payoff function}.
\end{definition}
\noindent
Here a \emph{play} of the game is a choice by each player of a particular strategy $s_i$ the collection of which forms a \emph{strategy profile} $(s_1, \cdots, s_n)$ whose corresponding \emph{outcome profile} is $G(s_1, \cdots, s_n)=(\omega_1, \cdots, \omega_n)$, where the $\omega_i$'s represent each player's individual outcome. Note that by assigning a real valued \emph{utility} to each player which quantifies that player's preferences over the various outcomes, we can without loss of generality, assume that the $\Omega_i$'s are all copies of $\mathbb{R}$, the field of real numbers.

In game theory one is frequently concerned with the identification of special strategies or strategic profiles. For example, most players would love to identify a strategy that guarantees a maximal utility. As this is not usually possible, a \emph{security strategy}, that is, a strategic choice that guarantees an explicit lower bound to the utility received, is also sought. But for a fixed $(n-1)$-tuple of opponents' strategies, rational players seek a \emph{best reply}, that is a strategy $s^{\star}_i \in S_i$ that delivers a utility at least as great, if not greater, than any other strategy $s_i \in S_i$. That is \[ G(\star, \cdots, \star, s^{\star}_i, \star, \cdots, \star) \geq G(\star, \cdots, \star, s_i, \star, \cdots, \star) \hspace{.5cm} \forall s_i \in S_i \]

A \emph{Nash equilibrium (NE)} for $G$ is a strategy profile $(s_1, s_2, ..., s_n)$ such that each $s_i$ is a best reply to the $(n-1)$-tuple of opponents' strategies. Other ways of expressing this concept include the observation that no player can increase his or her payoffs by unilaterally deviating from his or her equilibrium strategy, or that at equilibrium a player's opponents are indifferent to that player's strategic choice. As an example, consider the Prisoner's Dilemma, a two player game where each player has exactly two strategies (a so-called $2 \times 2$ or \emph{bimatrix} game) whose payoff function is indicated by the tableau below
\begin{table}[h]
	\centering
	  %Player One
	  \begin{tabular}{r|r|r|}
	    %\multicolumn{3}{c}{Player Two}\\
	    &$t_1$&$t_2$\\
		  \hline
      $s_1$&$(3,3)$&$(0,5)$\\
      \hline
			$s_2$&$(5,0)$&$(1,1)$\\
			\hline
	  \end{tabular}
	\caption{Prisoner's Dilemma}
	\label{tab:PrisonerSDilemma}
\end{table}
\noindent

Here, note that for player 1 the pure strategy $s_2$ always delivers a higher outcome than the strategy $s_1$ (say $s_2$ \emph{strongly dominates} $s_1$) and for player 2 the strategy $t_2$ strongly dominates $t_1$. Hence the pair $(s_2, t_2)$ is a (unique) Nash Equilibrium. However, games need not have equilibria amongst the pure strategy profiles as exemplified by the $2\times2$ game of Simplified Poker whose payoff function is given by

\begin{table}[h]
	\centering
	  %Player One
	  \begin{tabular}{r|r|r|}
	    %\multicolumn{3}{c}{Player Two}\\
	    &$t_1$&$t_2$\\
		  \hline
      $s_1$&$(5/4,-5/4)$&$(0,0)$\\
      \hline
			$s_2$&$(0,0)$&$(5/2,-5/2)$\\
			\hline
	  \end{tabular}
	\caption{Simplified Poker}
	\label{tab:SimplifiedPoker}
\end{table}

Classical game theoretic formalism now calls upon the theorist to extend the game $G$ by enlarging the domain and extending the payoff function. Of course, the question of if and how a given function extends is a time honored problem in mathematics and the careful application of the mathematics of extension is what will give our formalism of quantization its power. Returning to classical game theory, a standard extension at this point is to consider for each player the set of \emph{mixed strategies}, that is, the set of probability distributions over $S_i$. For a given set $X$, denote the probability distributions over $X$ by $\Delta(X)$ and note that when $X$ is finite, with $k$ elements say, the set $\Delta(X)$ is just the $k-1$ dimensional simplex $\Delta^(k-1)$ over $X$, i.e., the set of real convex linear combinations of elements of $X$. Of course, we can embed $X$ into $\Delta(X)$ by considering the element $x$ as mapped to the probability distribution which assigns 1 to $x$ and 0 to everything else. For a given game $G$, denote this embedding of $S_i$ into $\Delta(S_i)$ by $e_i$.

Now our game $G$ can be extended to a new, larger game $G^{mix}$, as follows. Given a profile $\left(p_1, \dots, p_n\right)$ of probability distributions over the $S_i$'s, by taking the product distribution we obtain a probability distribution over the product $\prod S_i$. Taking the push out by $G$ of this probability distribution we obtain a probability distribution over the image of $G$. By following this by the expectation operator we obtain the {\it expected outcome of the mixed strategy profile} $\left(p_1, \dots, p_n\right)$. Assigning the expected outcome to each mixed strategy profile we obtain the extended game 
$$
G^{mix}: \prod \Delta(S_i) \rightarrow \prod \Omega_i
$$

Note $G^{mix}$ is a true extension of $G$ as $G^{mix} \circ \Pi e_i = G$; that is we have the following commutative diagram
\clearpage
\begin{figure}[h]
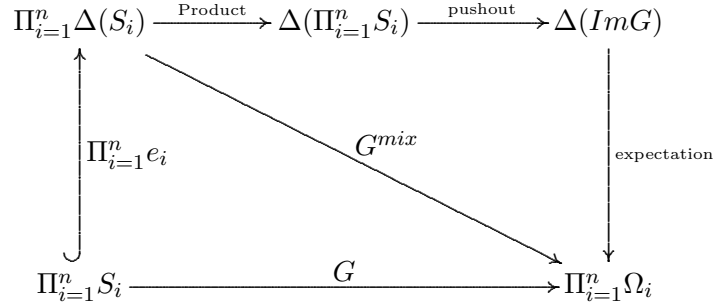

\[ \begindc{0}[50]
\obj(0,2){$\Pi_{i=1}^n \Delta(S_i)$}%{$\prod \Delta(S_i)$}
\obj(2,2){$\Delta (\Pi_{i=1}^n S_i)$}
\obj(4,2){$\Delta(ImG)$}%{$\Delta(ImG)$}
\obj(4,0){$\Pi_{i=1}^n \Omega_i$}%{$\prod \Omega_i$}
\obj(0,0){$\Pi_{i=1}^n S_i$}%{$\prod S_i$}
\mor{$\Delta (\Pi_{i=1}^n S_i)$}{$\Delta(ImG)$}{$\mbox{\tiny{pushout}}$}
\mor{$\Pi_{i=1}^n \Delta(S_i)$}{$\Delta (\Pi_{i=1}^n S_i)$}{$\mbox{\tiny{Product}}$}%{$\prod \Delta(S_i)$}{$\Delta(ImG)$$}{}
%\mor{$\Delta (\Pi_{i=1}^n S_i)$}{$\Delta(ImG)$}{}%{$\prod \Delta(S_i)$}{$\prod \Omega_i$}{$G^{mix}$}
\mor{$\Delta(ImG)$}{$\Pi_{i=1}^n \Omega_i$}{$\mbox{\tiny{expectation}}$}%{$\Delta(ImG)$}{$\prod \Omega_i$}{$\Sigma$}
\mor{$\Pi_{i=1}^n S_i$}{$\Pi_{i=1}^n \Delta(S_i)$}{$\Pi_{i=1}^n e_i$}[\atright, \injectionarrow]
\mor{$\Pi_{i=1}^n S_i$}{$\Pi_{i=1}^n \Omega_i$}{$G$}%{$\prod S_i$}{$\prod \Omega_i$}{$G$}
\mor{$\Pi_{i=1}^n \Delta(S_i)$}{$\Pi_{i=1}^n \Omega_i$}{$G^{mix}$}
\enddc \]	
	\caption{Extension of $G$ by $G^{mix}$}
	\label{fig:ExtensionOfGByGMix}
\end{figure}

Nash's famous theorem  \cite{binmore} says that if $S_i$ are all finite, then there always exists an equilibrium in $G^{mix}$. Unfortunately, this equilibrium is called a \emph{mixed strategy equilibrium for $G$}, when it is not an equilibrium of $G$ at all, the abusive terminology confusing $G$ with its image, Im$G$. Indeed, this is where much of the confusion in quantum game theory begins. 

But before proceeding onto quantization, it is useful to place other classical game theoretical ideas such as classical mediated communication and Aumann's notion of a \emph{correlated} equilibrium into this context. One begins by observing that the function from $\prod_{i=1}^n \Delta(S_i) \rightarrow \Delta(ImG)$ is not necessarily onto. As an example consider any $2 \times 2$ game $G$. If player 1 plays his first pure strategy with probability $p$, say, and player 2 plays her second pure strategy with probability $q$, say, the resulting probability distribution over the outcomes of $G$ is given by the tableau below: 

\[
	  %Player One
	  \begin{tabular}{r|r|r|}
	    %\multicolumn{3}{c}{Player Two}\\
	    &$t_1$&$t_2$\\
		  \hline
      $s_1$&$p(1-q)$&$pq$\\
      \hline
			$s_2$&$(1-p)(1-q)$&$(1-p)q$\\
			\hline
	  \end{tabular}
\]\\

\noindent
An easy exercise now shows that the element of $\Delta(ImG)$ represented by

\[
	  %Player One
	  \begin{tabular}{r|r|r|}
	    %\multicolumn{3}{c}{Player Two}\\
	    &$t_1$&$t_2$\\
		  \hline
      $s_1$&$1/2$&$0$\\
      \hline
			$s_2$&$0$&$1/2$\\
			\hline
	  \end{tabular}
\]\\

\noindent
is not realizable by any choice of $p$ and $q$. Classical mediated communication addresses this issue. Suppose during pre-play negotiation the players are able to hire a referee for negligible cost. For a given $\rho \in \Delta(ImG)$ the referee is meant to enforce $\rho$ as follows. The referee secretly observes a random event with probability distribution $\rho$, thus determining an outcome of $G$. The referee then communicates to each player only his or her strategic choice which yields the observed outcome.

Note that the players are no longer playing the game $G$, but in fact a much larger game $G_{\rho}^{com}$ which is easily described for $2 \times 2$ games and whose generalization to games with larger strategic spaces should be clear from our description. Suppose the strategic space for each player is represented by the pair $S=\left\{A, B\right\}$. The strategic spaces for $G_{\rho}^{com}$ can be represented by the quadruple $T=\left\{A',B',C',D'\right\}$ where the strategy $C'$ represents a player always cooperating with the referee, $D'$ represents the strategy where the player always deviates from the referee's instruction (i.e. playing $B$ when he hears $A$ and vice-versa), $A'$ represents cooperating with the referee when $A$ is recommended and deviating otherwise, and $B'$ represents cooperating with the referee when $B$ is recommended and deviating otherwise.

Two other things to note here. First, if both players choose to play $C'$, then the outcome of the new game is exactly the expected outcome under $\rho$. Second, $G_{\rho}^{com}$ extends the original game $G$ as there are embeddings $f_i:\left\{A, B\right\} \rightarrow \left\{A',B', C', D'\right\}$ taking $A$ to $A'$ and $B$ to $B'$ such that $G=G_{\rho}^{com} \circ \prod_{i=1}^2 f_i$, as in the diagram below

\begin{figure}[h]
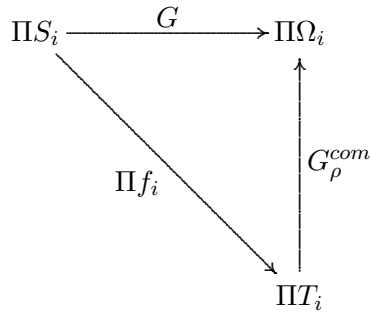

\[ \begindc{0}[50]
\obj(2,0){$\Pi T_i$}
\obj(2,2){$\Pi \Omega_i$}
\obj(0,2){$\Pi S_i$}
\mor{$\Pi S_i$}{$\Pi \Omega_i$}{$G$}
\mor{$\Pi T_i$}{$\Pi \Omega_i$}{$G_{\rho}^{com}$}[\atright, \solidarrow]
\mor{$\Pi S_i$}{$\Pi T_i$}{$\Pi f_i$}[\atright, \solidarrow]
\enddc \]	
	\caption{Extension of $G$ by $G_{\rho}^{com}$}
	\label{fig:ExtensionOfGByGCom}
\end{figure}

\noindent
Hence, classical mediated communication gives a \emph{family}, indexed by $\Delta(ImG)$, of extensions of $G$.

Following Aumann \cite{binmore}, a \emph{correlated equilibrium for} $G$ occurs whenever $(C', C')$ is a Nash equilibrium in $G_{\rho}^{com}$. That is, the players' agreement to follow the referee is \emph{self policing}, meaning that there is no gain to a player from unilateral deviating from the referee's recommendations. Note again the abusive terminology, the strategic choice for a correlated equilibrium is not a strategic choice for $G$ at all, but rather a strategic choice outside the embedded strategies for $G$ in a larger game. Of course, the use of correlated equilibrium may or may not improve the lot of the players. A classic example of correlated equilibrium improving the players' lot is given by the variant of the $2 \times 2$ game of Chicken given below
\begin{table}[h]
	\centering
	  %Player One
	  \begin{tabular}{r|r|r|}
	    %\multicolumn{3}{c}{Player Two}\\
	    &$t_1$&$t_2$\\
		  \hline
      $s_1$&$(2,2)$&$(0,3)$\\
      \hline
			$s_2$&$(3,0)$&$(-1,-1)$\\
			\hline
	  \end{tabular}
	\caption{Chicken}
	\label{tab:Chicken}
\end{table}

An easy exercise shows that $(s_2, t_1)$ and $(s_1, t_2)$ are both pure strategy equilibria and there is a unique mixed strategy equilibrium where every player plays each of his or her pure strategies with equal probability. This mixed strategy equilibrium pays out 1 to each player. It is also easy to see that even without a referee any real convex linear combination of these three outcomes forms a self policing agreement between the players. For example, the players could jointly observe a fair coin and agree to play the $(s_1, t_2)$ if it falls Heads and $(s_2, t_1)$ if it falls Tails. Note that the expected outcome of this agreement is $\left(\frac{3}{2}, \frac{3}{2}\right)$ which is better than the outcome $(1, 1)$ from the mixed strategy equilibrium. But even better and outside this region is the correlated equilibrium arising from the probability distribution $\frac{1}{3}(2,2)+\frac{1}{3}(0,3)+\frac{1}{3}(3,0)$ yielding the outcome $(\frac{5}{3},\frac{5}{3})$.

An example where mediated communication does not improve the lot of the players is given by Prisoner's Dilemma. One easily checks that due to the strong domination present in each player's strategy set, players always have an incentive to deviate from the referee's instruction if $\rho$ assigns a non-zero probability to any outcome other than the Nash equilibrium $(s_2, t_2)$. This domination is so strong that not even mixed strategy equilibrium that assign non-zero probability to an outcome other than the Nash equilibrium $(s_2, t_2)$ exists. A similar phenomenon occurs in the zero-sum game of Simplified Poker where any deviations from the equilibrium strategies where player I chooses his first strategy $\frac{2}{3}$ of the time and player II chooses her second strategy $\frac{1}{3}$ of the time is fully exploitable by the other player and hence an incentive to deviate from any other potential correlated equilibrium strategy. 

\section{A Formalism for Quantization}

Up until now, most authors have, like Meyer, focused their efforts on the quantization of the players' strategy spaces, essentially the domain of the payoff function that defines the game to be quantized (see for example \cite{Eisert, Gutoski, Landsburg, Meyer}). The principal technique is to identify these spaces with an orthogonal basis of some quantum system, in order that players may now take superpositions or even mixed superpositions of strategic choices by acting on the system via quantum operations. In addition, players may now correlate their strategic choices via the entanglement of the joint states of the system. Frequently, mere access to the higher randomization of superposition (as opposed to real probabilistic combination) or the correlation of strategic choices via entanglement allows payoffs to the players superior to those available in the game and its classical extensions. 

As a result, over the last decade, much discussion has occurred in the literature as to whether or not various protocols for achieving these happy results are indeed ``quantum'' or contain something new for game theory \cite{Benjamin, Van Enk}. These discussion have focused on the situations when not all players have access to quantized strategy spaces, as in Meyer's original penny flip game \cite{Meyer}, classical methods for realizing the correlation given by entanglement exist as for the MW and EWL protocols \cite{Van Enk} or as in the Duetchse-Josza algorithm \cite{DJ}, and the games based on the conditional probability distributions found in Bell's theorem as introduced by Landsburg and Dahl \cite{Dahl} for classical duopoly, and by Iqbal and Cheon for Prisoner's Dilemma, Chicken, Stag Hunt, and Battle of the Sexes \cite{Iqbal}. 

The main assertion in \cite{Bleiler} is that all of these controversies resolve via the use of a mathematical formalism for game quantization that focuses on the quantization of the {\it payouts} of the original game $G$ to be quantized, and expresses the quantized version as a (proper) extension of the original payout function in the set-theoretic sense, just as in the classical case where exhaustive studies are made of the extensions of classical games obtained from the probabilistic combinations of payoffs. As mentioned above, these combinations arise by the players mixing their strategic choices or by communicating their strategic choices to a referee. The protocol also allows for fresh quantum game theoretic interpretations of a number of quantum algorithms and for a broad swath of operations in quantum logic synthesis \cite{Bleiler-Perkowski, Khan}. 

Classically, we constructed the probability distributions over the outcomes of a game $G$. We now wish to pass to a more general notion of randomization, that of quantum superposition. Begin then with a Hilbert space $\mathcal{H}$ that is a complex vector space equipped with an inner product. For the purpose here assume that $\mathcal{H}$ is finite dimensional, and that we have a finite set $X$ which is in one-to-one correspondence with an orthogonal basis $\mathcal{B}$ of $\mathcal{H}$. 

By a {\it quantum superposition} of $X$ with respect to the basis $\mathcal{B}$ we mean a complex projective linear combination of elements of $X$; that is, a representative of an equivalence class of complex linear combinations where the equivalence between combinations is given by non-zero scalar multiplication. Quantum mechanics calls this scalar a {\it phase}. When the context is clear as to the basis to which the set $X$ is identified, denote the set of quantum superpositions for $X$ as $QS(X)$. Of course, it is also possible to define quantum superpositions for infinite sets, but for the purpose here, one need not be so general. What follows can be easily generalized to the infinite case. 

As the underlying space of complex linear combinations is a Hilbert space, we can assign a length to each linear combination and, up to phase, always represent a projective linear combination by a complex linear combination of length 1. This process is called {\it normalization} and is frequently useful. 

For each quantum superposition of $X$ we can obtain a probability distribution over $X$ by assigning to each component the ratio of the square of the length of its coefficient to the square of the length of the combination. For example, the probability distribution produced from $\alpha x + \beta y$ is just
$$
\frac{\left|\alpha\right|^2}{\left|\alpha\right|^2+\left|\beta\right|^2}x+\frac{\left|\beta\right|^2}{\left|\alpha\right|^2+\left|\beta\right|^2}y
$$
Call this function $QS(X) \rightarrow \Delta(X)$ a {\it quantum measurement with respect to $X$}, and note that geometrically quantum measurement is defined by projecting a normalized quantum superposition onto the various elements of the normalized basis $\mathcal{B}$. Denote this function by $q_X^{meas}$, or if the set $X$ is clear from the context, by $q^{meas}$. 

Now given a finite $n$-player game $G$, suppose we have a collection $Q_1, \dots, Q_n$ of non-empty sets and a \emph{protocol}, that is, a function $\cal{Q}$ $:\prod Q_i \rightarrow QS(\rm{Im}G)$. Quantum measurement $q_{\rm{Im}G}^{meas}$ then gives a probability distribution over $\rm{Im}G$. Just as in the mixed strategy case we can then form a new game $G^{\cal Q}$ by applying the expectation operator. Call the game $G^{\cal Q}$ thus defined to be the {\it quantization of $G$ by the protocol $\mathcal{Q}$}. Call the $Q_i$'s sets of {\it pure quantum strategies} for $G^{\cal Q}$. Moreover, if there exist embeddings $e'_i:S_i \rightarrow Q_i$ such that $G^{\cal Q} \circ \prod e'_i=G$, call $G^{\cal Q}$ a {\it proper} quantization of $G$. If there exist embeddings $e''_i:\Delta(S_i) \rightarrow Q_i$ such that $G^{\cal Q} \circ \prod e''_i=G^{mix}$, call $G^{\cal Q}$ a {\it complete} quantization of $G$. These definitions are summed up in the following commutative diagram. Note for proper quantizations the original game is obtained by restricting the quantization to the image of $\prod e'_i$. For general extensions, the Game Theory literature refers to this as ``recovering'' the game $G$. 

\begin{figure}[h]
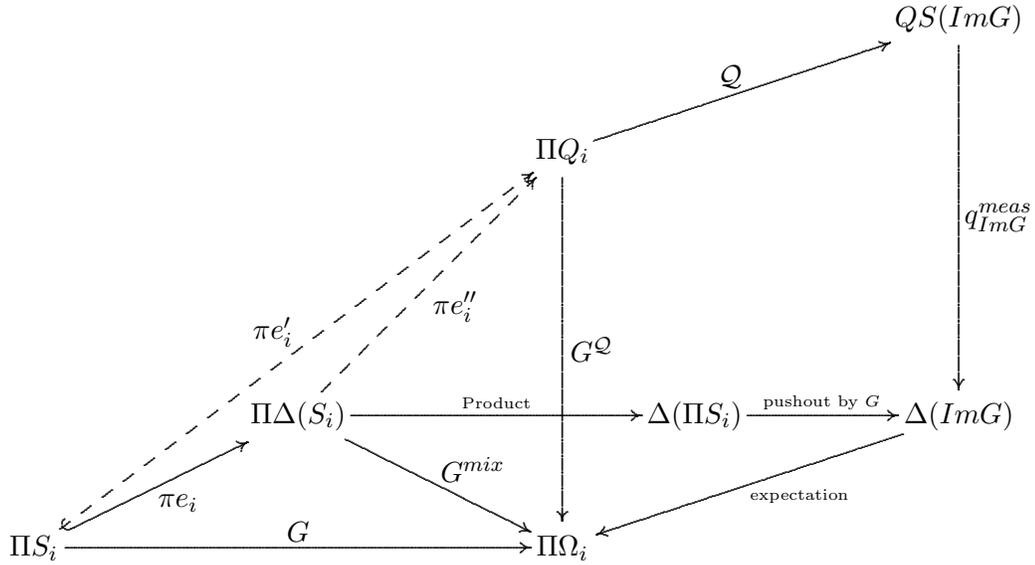

	\[ \begindc{0}[50]
\obj(4,3){$\Pi Q_i$}
\obj(7,4){$QS(ImG)$}
\obj(7,1){$\Delta(ImG)$}
\obj(5,1){$\Delta (\Pi S_i)$}
\obj(2,1){$\Pi \Delta(S_i)$}
\obj(4,0){$\Pi \Omega_i$}
\obj(0,0){$\Pi S_i$}
\mor{$\Pi Q_i$}{$QS(ImG)$}{$\cal Q$}
\mor{$\Pi Q_i$}{$\Pi \Omega_i$}{$G^{\cal Q}$}[\atleft, \solidarrow]
\mor{$QS(ImG)$}{$\Delta(ImG)$}{$q_{ImG}^{meas}$}
\mor{$\Delta(ImG)$}{$\Pi \Omega_i$}{$\mbox{\tiny{expectation}}$}
\mor{$\Pi S_i$}{$\Pi \Omega_i$}{$G$}
\mor{$\Pi \Delta(S_i)$}{$\Pi Q_i$}{$\pi e^{\prime \prime}_i$}[\atright, \dasharrow]
\mor{$\Pi S_i$}{$\Pi Q_i$}{$\pi e^{\prime}_i$}[\atleft, \dasharrow]
\mor{$\Pi S_i$}{$\Pi \Delta(S_i)$}{$\pi e_i$}[\atright, \injectionarrow]
\mor{$\Pi \Delta(S_i)$}{$\Delta (\Pi S_i)$}{$\mbox{\tiny{Product}}$}%[\atleft, \dasharrow]
\mor{$\Delta (\Pi S_i)$}{$\Delta(ImG)$}{$\mbox{\tiny{pushout by $G$}}$}
\mor{$\Pi \Delta(S_i)$}{$\Pi \Omega_i$}{$G^{mix}$}
\enddc \]
	\caption{A Quantization Formalism}
	\label{fig:ACompleteQuantizationOfG}
\end{figure}

Note that the definitions of $G^{mix}$ and $G^{\cal Q}$ show that a complete quantization is proper. Furthermore, note that finding a mathematically proper quantization of a game $G$ is now just a typical problem of extending a function. It is also worth noting here that nothing prohibits us from having a quantized game $G^{\cal Q}$ play the role of $G$ in the classical situation and by considering the probability distributions over the $Q_i$, creating a yet larger game $G^{m \cal{Q}}$, the {\it mixed quantization of G with respect to the protocol $\cal{Q}$}. For a proper quantization of $G$, $G^{m \cal{Q}}$ is an even larger extension of $G$. The game $G^{m \cal{Q}}$ is described in the commutative diagram of Figure \ref{fig:ExtensionOfGByGQm}.

\begin{figure}[h]
\[ \begindc{0}[50]
\obj(0,2){$\Pi_{i=1}^n \Delta(Q_i)$}%{$\prod \Delta(S_i)$}
\obj(2,2){$\Delta (\Pi_{i=1}^n Q_i)$}
\obj(5,2){$\Delta(ImG^{Q})$}%{$\Delta(ImG)$}\coprod
\obj(5,0){$\Pi_{i=1}^n \Omega_i$}%{$\prod \Omega_i$}
\obj(0,0){$\Pi_{i=1}^n Q_i$}%{$\prod S_i$}
\mor{$\Delta (\Pi_{i=1}^n Q_i)$}{$\Delta(ImG^{Q})$}{$\mbox{\tiny{pushout by $G^Q$}}$}
\mor{$\Pi_{i=1}^n \Delta(Q_i)$}{$\Delta (\Pi_{i=1}^n Q_i)$}{$\mbox{\tiny{Product}}$}%{$\prod \Delta(S_i)$}{$\Delta(ImG)$$}{}
%\mor{$\Delta (\Pi_{i=1}^n S_i)$}{$\Delta(ImG)$}{}%{$\prod \Delta(S_i)$}{$\prod \Omega_i$}{$G^{mix}$}
\mor{$\Delta(ImG^{Q})$}{$\Pi_{i=1}^n \Omega_i$}{$\mbox{\tiny{expectation}}$}%{$\Delta(ImG)$}{$\prod \Omega_i$}{$\Sigma$}
\mor{$\Pi_{i=1}^n Q_i$}{$\Pi_{i=1}^n \Delta(Q_i)$}{$\pi \tilde{e}_i$}[\atright, \injectionarrow]
\mor{$\Pi_{i=1}^n Q_i$}{$\Pi_{i=1}^n \Omega_i$}{$G^{\cal Q}$}%{$\prod S_i$}{$\prod \Omega_i$}{$G$}
\mor{$\Pi_{i=1}^n \Delta(Q_i)$}{$\Pi_{i=1}^n \Omega_i$}{$G^{m \cal{Q}}$}
\enddc \]	
	\caption{Extension of $G$ by $G^{m \cal{Q}}$}
	\label{fig:ExtensionOfGByGQm}
\end{figure}

Note that the quantum strategy sets $Q_i$ need not consist of quantum superpositions, although in many quantization protocols they do, see for example \cite{Eisert, Landsburg}. Indeed, protocols with classical inputs yielding quantum superpositions of the outcomes of certain games have already been posited \cite{Dahl, Iqbal}. These and some other specific protocols are discussed in the context of the formalism above in \cite{Bleiler}.

As discussed in \cite{Bleiler} and in part the following sections, the literature gives several protocols for quantizing one, two, and occasionally even multi-player games, some improper, some proper but not complete, and some yielding complete quantizations. Yet there is an ongoing debate in the literature as to what is the `correct' method of quantizing a game. The above formalism suggests that this is just the wrong question to ask, as under this formalism a given game can admit several different quantizations. It also makes clear that comparisons between various quantizations, between quantizations and various classical extensions, and between quantizations and the original game itself often amounts to comparing ``apples'' to ``oranges''. A specific example appears next.

\section{Communication via the EWL Protocol}

In classical mediated communication, players have a referee mediate their game and the communication of their strategic choices. For simplicity, assume our players have but two classical pure strategies to choose from. The communication of each players strategic choices is implemented by the sending of bits to the players, put into an initial state by the referee. Presumably players then send back their individual bits in the other state (Flipped) or in the original state (Un-Flipped) to indicate the choice of their second or first classical pure strategy respectively. The bits are then examined by the referee who then makes the appropriate payoffs. 

When the communication between the referee and the players is over quantum channels, Eisert, Wilkens and Lewenstein \cite{Eisert} have proposed families of quantization protocols that depend on the initial joint state prepared by the referee. Players and the referee communicate via {\it qubits}, a two pure state quantum system with a fixed observational basis.  In the EWL protocol the referee determines the outcomes via a new observational basis corresponding to the actions of (No Flip, No Flip), (No Flip, Flip), (Flip, No Flip), (Flip, Flip) by the players. Players may choose from any physical operation (i.e. the Lie group $S(2)$) as pure quantum strategies (the $Q_i$'s in the formalism above) or even probabilistic combinations thereof (the $\Delta Q_i$'s in the formalism) for their strategic choices. The procedure above describes for each initial state $I$ a protocol ${\cal{Q}}_I$ and a quantized and mixed quantized game $G^{{\cal{Q}}_I}$ and $G^{{m\cal{Q}}_I}$ per the formalism. 

If the initial state prepared by the referee is given in the Dirac notation by $\left|0\right\rangle \otimes \left|0\right\rangle$, then the EWL protocol is a complete quantization and is in fact equivalent to the classical game $G^{mix}$. But when the initial state is given by the maximally entangled state $\left|0\right\rangle \otimes \left|0\right\rangle+\left|1\right\rangle \otimes \left|1\right\rangle$, the EWL protocol remains complete and sets up an onto map from the product of the strategy spaces to $\Delta\left(ImG\right)$.

It is here that a fundamental question arises; that is, are Nash equilibria in such a quantized games truly new? That is, is the probability distribution that arises from an equilibrium pair in the quantized version of game $G$ different from that arising from a classical correlated equilibrium for $G$? The maximally entangled EWL quantization described above admits a mixed quantum strategy equilibrium where each player uses the uniform probability distribution over his or her choice of pure quantum strategy \cite{Landsburg}. The resulting probability distribution over the payoffs of $G$ is now again the uniform distribution, assigning an equal probability to each of the four outcomes of $G$. Note that for the Prisoner's Dilemma, this distribution does {\it not} arise from a classical correlated equilibrium as it assigns a non-zero probability to each of the classical non-equilibrium payoffs, and so does not correspond to any classical correlated equilium for this game, yet delivering a payoff to the players superior to that of the classical pure strategy equilibrium. An even more remarkable result holds true for the maximally entangled EWL quantization of the zero-sum game of Simplified Poker, where the uniformly mixed quantum equilibrium out performs the classical mixed strategy equilibrium payoff for player I, yet is still a security strategy for player I against which player II has no recourse. 

\section{Conclusion}

The discussion above shows that for the Prisoners' Dilemma and Simplified Poker quantization does indeed hold something new for Game Theory.  Several other controversies can be resolved and new game theoretic interpretations of certain problems from quantum computation and quantum logic synthesis can be illuminated via the quantization formalism described here, see \cite{Bleiler, Bleiler-Perkowski, Khan} for details.


\begin{thebibliography}{99}

\bibitem{Benjamin}
S. C. Benjamin,
{\em Comment on: ``A quantum approach to static games of complete information"},Phys. Lett. A 272 (2000) 291.

\bibitem{binmore}
K.~Binmore,
{\it Fun and Games: A Text on Game Theory}, D.C. Heath (October 1991).

	\bibitem{Bleiler} 
	S. A. Bleiler,
{\em A Formalism for Quantum Games}, preprint, Portland State University.

\bibitem{Bleiler-Perkowski}
S. A. Bleiler, M. Perkowski,
Portland State University, preprint.

\bibitem{Dahl}
G. Dahl, S. Landsburg,
{\em Quantum Strategies in Non cooperative Games}, University of Rochester, preprint.

\bibitem{DJ}
D. Deutsch, R. Jozsa,
{\em Rapid Solutions of Problems by Quantum Computation}, Proceedings of the Royal Society of London A 439: 553.
	
	\bibitem{Eisert}
J. Eisert, M. Wilkens, M. Lewenstein
{\em Quantum Games and Quantum Strategies}, Physical Review Letters, 
Vol. 83, Number 15, October 11, 1999.

\bibitem{Gutoski}
G. Gutoski, J. Watrous,
{\em Toward a general theory of quantum games}, Proceedings of the thirty-ninth annual ACM symposium on Theory of computing, 2007.
 
 \bibitem{Iqbal}
  Azhar Iqbal, Taksu Cheon,
 {\em Constructing quantum games from nonfactorizable joint probabilities}, Phys. Rev. E 76, 061122 (2007).

\bibitem{Khan}
F. S. Khan,
Portland State University, preprint. 

\bibitem{Landsburg}
S. Landsburg,
{\em Nash Equilibria in Quantum Games,} University of Rochester - Center for Economic Research 
(RCER) Working Paper Number 524.

%\bibitem{Marinatto}
%   L. Marinatto, T. Weber,
%    {\em A Quantum Approach to Static Games of Complete Information}, Phys. Lett. A 271, 291 (2000).   
	
	\bibitem{Meyer}
	  D. A. Meyer, \emph{Quantum Strategies}, Phys. Rev. Lett. $\mathbf{82}$, 1052-1055, 1999.
	  
	  \bibitem{Van Enk}
S. J. van Enk, R. Pike,
{\em Classical rules in quantum games}
Physical Review A 66, 024306 2002.

\end{thebibliography}
\end{document}